# Thermal diffusivity of PMMA/Alumina Nano Composites Using Molecular Dynamic Simulation

Maryam Mohammadi[a, *], Jamal Davoodi[a]

[a,] Department of Physics, University of Zanjan, Zanjan 45195-313, Iran

*Corresponding Author.
E-mail address: mohamady_maryam@znu.ac.ir

## Abstract

Fire protection mainly takes place through physical or chemical pathways or both of them. The thermal diffusivity is one of the basic thermophysical properties which can connect the chemical structure. It is well-known in the literature that there exist relationships between the thermal diffusivity and the thermal stability. Also, the thermal diffusivity can be related to some fire retardant properties such as total-heat-release (THR), time-to-ignition (TTI) and peak-heat-rare-release (PHRR) that are of the most important parameters in the evaluation of potential fire hazard of a given material. Metal oxides, as one of the most promising flame retardant additives, improve the fire-retardant and the thermal stability properties of polymers. In the present study, molecular dynamic (MD) simulations based on the united atom model are utilized to study the effect of alumina nanoparticles on the thermal diffusivity of isotactic Polymethyl methacrylate (is-PMMA) polymer. Thermal diffusivity of PMMA and PMMA/alumina nanocomposite were investigated through calculating heat capacity, density and thermal conductivity in the range of 300-700 K. Heat capacity can be calculated using fluctuations properties. Thermal conductivity was calculated through the nonequilibrium MD (NEMD) simulation by means of the Fourier's law approach. Results show that the alumina nanoparticles decrease the amount of heat capacity and increase the glass transition temperature ($T_g$), thermal conductivity and thermal diffusivity of the PMMA.

**Highlight:**

1- The Heat capacity of PMMA and PMMA/alumina nanocomposite was calculated

2- The Thermal conductivity of PMMA versus temperature was calculated.

3- The Thermal diffusivity was investigated.

# 1 Introduction

Fire annually kills thousands of people and causes heavy damages all over the world. For instance, 400,000 home fires occur every year in the United States, causing approximately 7 billion US dollars of damage [1, 2] .Thus, the improvement of fire retardancy of materials is a major concern. In order to control and reduce the detrimental effects of the fire, a comprehensive study on the fire behavior and mechanisms is essential. Considerable efforts have been carried out for a better understanding of the involving mechanisms and behaviors. In general, physical or chemical pathways or both are involved in flame retardancy mechanisms. The key factors influencing the fire behavior are chemical such as thermal stability or/and physical such as thermal diffusivity where addressed. Among them, the thermal diffusivity has a key role in the prediction of the flammability property.

The thermal diffusivity is the ability of a material to transmit the heat rather than to absorb it. Consequently, the higher thermal diffusivity, the lower time is required to pass the heat through the material. As a result, the thermal diffusivity can physically affect the rate at which a given material decomposes [3]. It is recently found that increasing the thermal diffusivity mostly increases the time-to-ignition (TTI) while decreases both the peak-heat-rare-release (PHRR) and the total-heat-release (THR). The PHRR and THR are found as the two most essential parameters of a given material in the evaluation of its potential fire hazard [3]. As a result, when thermal diffusivity is increased the flammability is reduced [3].

Polymer and polymer products have extensive applications ranging from the household appliance, computers, furniture, to many special industrial parts. Polymers have the advantages such as good processability, light weight, low water absorption, high electrical resistivity, high voltage breakdown, strength, corrosion resistance, and most importantly low cost. However, the flammability of polymers has been regarded as a key characteristic and a challenging concern. Hence, the improvement of the fire retardancy of polymer materials is highly desirable when it comes to the polymers fire resistance properties. It is necessary, therefore, to find a solution in order to limit the risks through the design of novel advanced materials with enhanced flammability behavior. Recently, polymer nanocomposite with the capability of being engineered as a sort of fire retardant materials have been increasingly developed and used in the manufacturing of many household and industrial products. Over the past decades, a variety of micro and nano fillers has been shown to be efficient additives utilized for the improvement of the polymer fire retardancy [4].

Recent studies show that the reinforcements such as metal oxides can decrease the flammability and enhance the thermal stability of polymers [3, 5-7]. As reported in the previous studies, the dominant mechanism of fire retardancy of a metal oxide is mostly attributed to its physical properties rather than chemical ones [8]. For instance, metal dioxide nanoparticles such as titanium dioxide, alumina and etc. are significantly able to modify the thermal degradation of a polymer [3, 5-7].

In this study, we focus on the examination of poly methyl methacrylate (PMMA) as the host matrix. The PMMA is a thermoplastic polymer which is widely used in various applications for its many advantageous properties. However, its major drawback is its flammability issues with its limiting oxygen index (LOI) of 18). To understand how metal oxides fillers improve the flammability of the PMMA, a literature review is presented in the following.

Researchers have widely investigated the effects of certain material variables on the thermal stability and flammability to obtain a better insight into the factors influencing polymer-metal oxide nano composite properties. Friederich et al. have examined alumina and titanium dioxide and bohemit fillers at various loading contents. It has been shown that the TTI is increased by approximately 25% at 15 wt% of the fillers irrespective of the reinforcement used. Also, the same study represents that the PHR is reduced by approximately 50% at 15 wt% of the same fillers. The THR is effectively reduced (approximately 30%) when alumina at 15 wt% is used. [3] .Laachachi has evaluated the effect of titanium dioxide and iron oxide fillers, revealing that both the fillers improve the thermal stability by about 70ºC and decrease the PHR [5]. The PHR rate is decreased and the total burn time is increased with increasing the percentage of both the fillers [8].

Molecular dynamic (MD) simulation is a powerful tool to study the behavior of polymers. Although there are some MD simulations on thermal decomposition and thermal degradation for polyethylene (PE) [9, 10], polypropylene (PP) and poly(isobutylene) (PIB) [11], poly-a-methyl styrene [12], amorphous polylactide [13], polyimide [14], (PE, PP, PS, PET, N6, PVDF) [15], PS and PET[16], to our knowledge there are no similar calculations for thermal diffusivity via the MD simulations. Through the present study, we intend to investigate how adding metal oxides can improve the fire retardancy of a polymer using MD simulations. For this purpose, we calculate the thermal diffusivities of the PMMA and the PMMA/Alumina nanocomposite and study how the alumina nanoparticles affect the thermal diffusivity of the PMMA. The thermal diffusivity can be calculated by [17]:

$$\alpha = \frac{\kappa}{C_p \rho} \qquad (1)$$

where $\kappa$ is the thermal conductivity, Cp is the heat capacity at a constant pressure, ρ is density and α is the thermal diffusivity. Therefore, in this paper the thermal diffusivity can be achieved by calculating the thermal conductivity, heat capacity and density.

## 2 Materials and MD Simulation Setup

All molecular simulations were done using the MD simulation package, LAMMPS (large-scale atomic/molecular massively parallel simulator), developed at the Sandia National Laboratories [18]. The system consisted of an alumina nanoparticle with the radius of 7 angstroms at the center of the simulation cubic and three linear isotactic PMMA (is-PMMA) chains with a degree of polymerization of 100 around the nanoparticle (10%weight alumina). The structure of the PMMA monomer was proposed by Shaffer et al. [19]. The data of alumina structure is taken from Mincryst [20], the crystallographic database from the Institute of Experimental Mineralogy of the Russian Academy of Sciences. The PMMA conformations using a random and self-avoiding walk algorithm and the alumina structure were generated in the Fortran 90 and then were imported into the LAMMPS as the data input. The chain structure of the PMMA and the alumina nanoparticle generated as described are shown in Figure 1. To perform the simulations, a three-dimensional periodic boundary condition was applied. To reduce the computation time, a united atom model (where each carbon is grouped with its bonded hydrogen atoms) was applied. An interatomic force field proposed by Okada et al. was used for modeling of the atomic interactions of the polymer chains [21]. Some modifications have been made to the force field equations to make them usable in the LAMMPS. Here, the interaction potential ( $U$ ) is expressed as the sum of the bonding and nonbonding interactions and is defined as the following expression

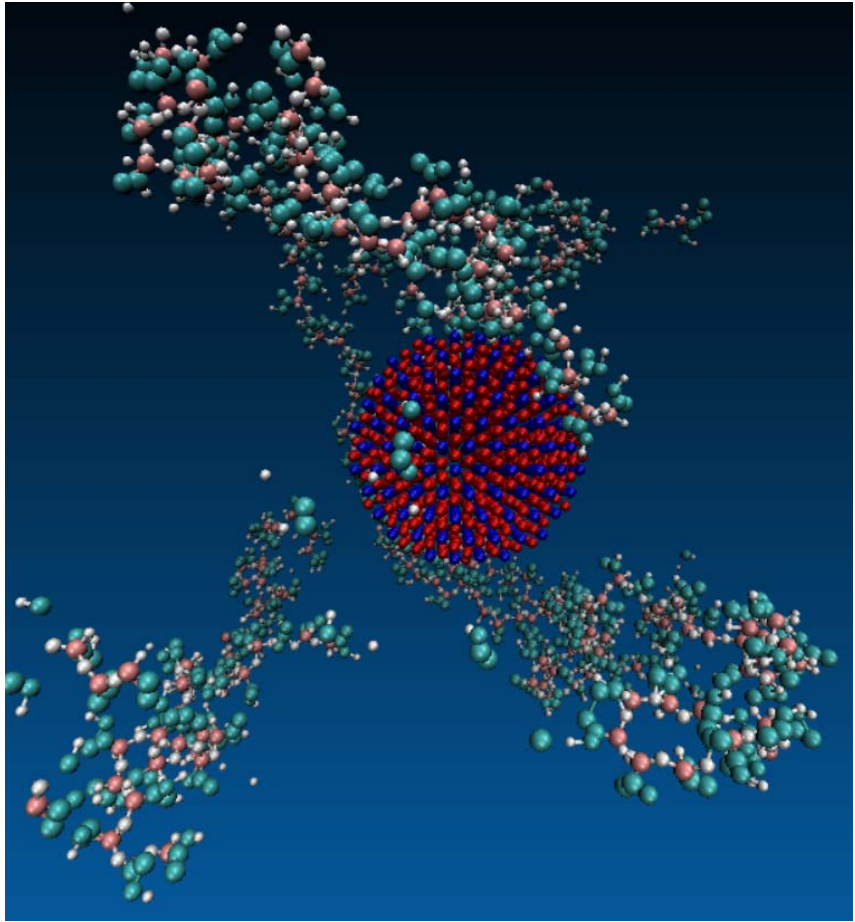
Figure 1. is-PMMA/alumina nanocomposite

$$U = \sum_{bonds} k_r(r-r_0)^2 + \sum_{angles} k_\theta(\theta-\theta_0)^2 + \sum_{torsions} \sum_{i=1} (V_n \cos n\phi)$$
$$+ \sum_{\substack{improper \\ torsions}} (K_1(\Theta-\Theta_0) + K_2(\Theta-\Theta_0)^2) + \sum \frac{A}{r^{12}} - \frac{C}{r^6} \quad (1)$$

where the first term is related the bond stretching energy, $r$ and $r_0$ are the bond length and the equilibrium length of the bond, respectively. The second term illustrates the angular bending energy, $\theta$ and $\theta_0$ are the bending angle and the equilibrium angle of the bond, respectively. The third and fourth terms correspond to the dihedral torsion and the improper torsion energies, respectively. $\phi$ is the dihedral torsion angle and $\Theta$ is the sum of the three neighboring bending angles, $\Theta_0$ is the equilibrium sum of the three neighboring bending angles. The last term is the Lennard–Jones energy between two non-bonded atoms/molecules.

The interaction potential between the PMMA and the alumina in the PMMA/alumina nanocomposite consisted of binding and non-binding potentials [22]. The binding potential defines the interaction between carbonyl and ester oxygen of the PMMA and the aluminum. The non-binding potential describes the interaction between carbons in the PMMA and the aluminum. The binding (Ub) and non-binding (Unb) potentials are expressed as

$$U_b = A\exp(-2\lambda(z_i - z_b)) - B\exp(-\lambda(z_i - z_b)) + C\lambda^2(z_i - z_b)^2 \exp(-\lambda(z_i - z_b))$$

$$U_{nb} = \frac{2}{3}\pi\sigma_i\varepsilon_i\rho\left[\frac{2}{15}\left(\frac{\sigma_i}{z_i}\right)^9 - \left(\frac{\sigma_i}{z_i}\right)^3 + \frac{2}{3}\sqrt{\frac{5}{2}}\right] \quad (2)$$

where $\rho$ is the number density of aluminum atom, $z_i$ indicates the distance of atom i from the aluminum surface, $\varepsilon_i$ and $\sigma_i$ are Lennard-Jones parameters between the non-binding atoms of the PMMA and the aluminum atoms in the substrate. The first two terms in Equ. (2) are the standard Morse potential and the third term represents the activation energy.

It is notable that since suggested potential by Shaffer and Chakraborty were written for united atom PMMA, there were no potential interactions with hydrogen in PMMA. The potential was expanded to alumina by assuming the interaction of the oxygen atoms in the alumina with PMMA could be described as the average non-binding potential interaction of the aluminum with the PMMA [22]. The First Principles Study on Polymer-Metal-Oxide Adhesion Was done by Drabold et al. Their result shows the oxygen atoms in alumina have very weak interactions to PMMA and the aluminum atoms interactions with the carbon and hydrogen in PMMA is weak too [23].

The suggested interatomic interactions for the alumina potential proposed by Streitz and Mintmire [24] define a many-body potential consisting of an embedded-atom (VEAM) and an electrostatic (VES) part as fallows

$$V_{EAM} = \sum_i F_i[\rho_i] + \sum_{i<j} \varphi_{ij}(r_{ij}) \quad (3)$$

$$\rho_i(r) = \sum_{i \neq j} \xi_j e^{-\beta(r_{ij} - r_j^*)}$$

$$F_i(\rho_i) = -A_i\sqrt{\frac{\rho_i}{\xi_i}}$$

$$\varphi_{ij}(r) = 2B_{ij}e^{\frac{\beta_{ij}}{2}(r-r_{ij}^*)} - C_{ij}\left[1+\alpha(r-r_{ij}^*)\right]e^{-\alpha(r-r_{ij}^*)}$$

Here, $F_i(\rho_i)$ represents the energy required to embed atom $i$ in an environment with an electron density $\rho_i$ and $\varphi_{ij}(r)$ is the pair-wise interaction.

The pair potential $\varphi_{ij}(r)$ becomes strongly repulsive at a small interatomic distance. The electrostatic part accounts for electric charges on atoms and is defined as

$$V_{ES} = \sum_i v_i(q_i) + \frac{1}{2}v_{ij}(r_{ij};q_i;q_j) \qquad (4)$$

$$v_i(q_i) = v_i(0) + \chi_i^0 q_i + \frac{1}{2}J_i^0 q_i^2$$

$$v_{ij}(r_{ij};q_i;q_j) = \int d^3r_1 \int d^3r_2 \rho_i(r_1;q_i)\rho_j(r_2;q_j)/r_{12}$$

## 3 Equilibrium

The three polymer chains were first built up around a spherical cavity with the same diameter of an alumina nanoparticle (For a neat polymer, the diameter was considered zero) in the simulation box at the temperature of T=700K. The polymer chains could move into the empty space where the nanoparticle would be inserted. To avoid this, a wall potential was placed around the space. The wall potential was a Leonard Jones one which parameters were derived from the average Leonard Jones parameters of Okada et al.'s PMMA potential [21]. To minimize the total potential energy of the initial system, the conjugate gradient method was applied.

This was followed by allowing the polymer chains to equilibrate. The equilibration of the polymer has been explained in previous work[25]. Next, the alumina nanoparticle was inserted into the cavity and then the system reached the equilibrium with the NPT ensemble after 2 ns. Then the system was cooled from 700 K to 300 K at a rate of 10K/5ns.

# 4 Results and discussion

## 4.1 Density

Figure 2. illustrates the density ($\rho$) of the PMMA system (neat polymer) and the PMMA/alumina nanocomposite versus a temperature ramp. An obvious change in the slope of the curve is observed. The temperature at which the slope of the curve undergoes a sharp variation is regarded as the glass transition temperature ($T_g$). At $T_g$, the physical properties of a polymer changes and the polymer transfers from a rubbery state (above $T_g$) to a glassy state (below $T_g$). According to Figure 2. both the density and the $T_g$ of the PMMA/alumina nanocomposite are larger than those of the PMMA. Figure 2 shows an increase of about 10 K in $T_g$ for the PMMA/alumina nanocomposite compared to that for the PMMA.

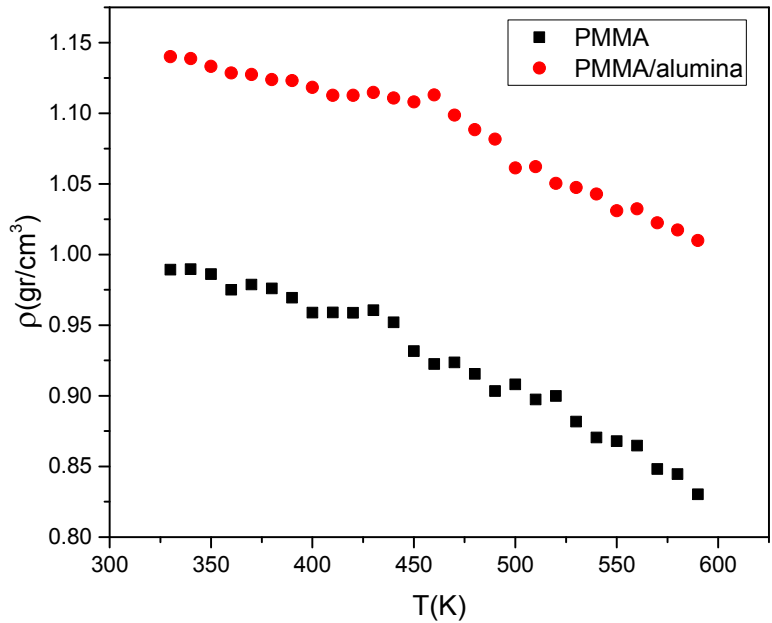

Figure 2. Density of (a) the PMMA and (b) the PMMA/alumina against temperature.

A polymer consists of two phases; one is the polymer chains and the other is the free volume called space phase. According to Figure 2, the density-temperature curve of the PMMA/alumina shows a smaller gradient than that of a pure polymer system, which could be due to a smaller free volume available for the motion of the polymer chains.

## 4.2 Thermal conductivity

There are two methods for the computation of the thermal conductivity (TC) based on the molecular dynamics which are the equilibrium MD (EMD) and the non-equilibrium MD (NEMD) approaches. In this study, the EMD simulation by means of the Fourier's law (**Error! Reference source not found.**6) was employed to investigate the thermal conductivity of the PMMA and alumina/PMMA nanocomposite.

$$\kappa = -\frac{J_q}{dT/dy} \quad (6)$$

where $\kappa$ is the thermal conductivity, $dT/dy$ is the steady-state temperature gradient in unidirectional heat flow inside of the unit cell and $J_q$ is the heat flux that is given with the following expression:

$$J_q = \frac{E}{2A_{cross-section}} \quad (7)$$

The simulation system was divided into several slabs on the total length, two slab at the two end will be the hot region and at the other end will be the cold region. Then the linear temperature region using the least square method was fitted to obtain the temperature gradient. The thermal conductivity κ can be calculated by **Error! Reference source not found.**6.

Figure 3 shows the TC versus temperature for the neat PMMA and the PMMA/alumina nanocomposite. It is found that with increasing the temperature up to the Tg, the thermal conductivity of the PMMA increases, while it decreases above Tg. This is due to the fact that at temperatures above the Tg, the free-volume increases and air has a low thermal conductivity. Such a temperature-dependent behavior of the thermal conductivity is in a good agreement with the experiments [26].

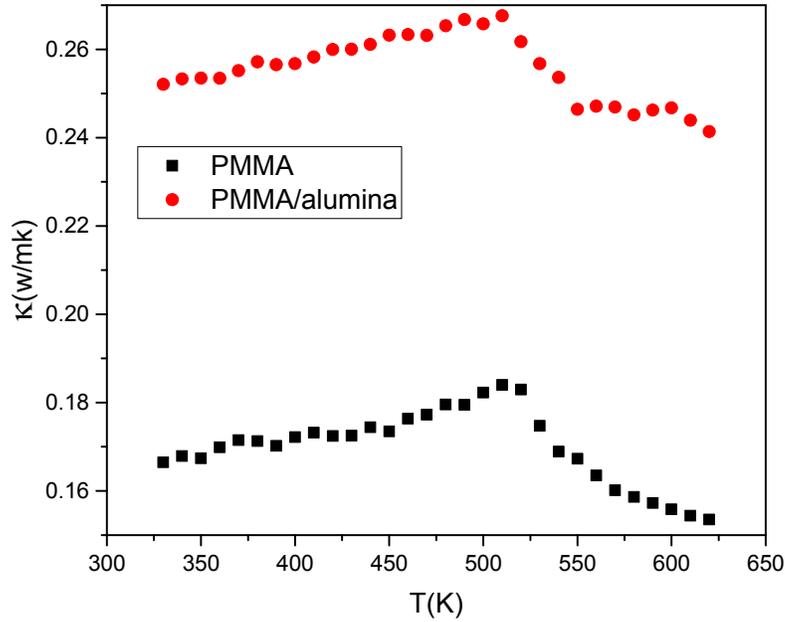

Figure 3. Thermal conductivity of (a) the PMMA and (b) the PMMA/alumina vs. temperature.

Although the thermal conductivity along a single chain is very high [27, 28], thermal conductivity in polymer bulk is very low due to the phonon scattered at and the thermal contact resistance ends of chains. Other limiting factors are random orientation, entanglement chains and voids which act as scattering phonon sites for the heat. The comparison between the PMMA and the PMMA/Alumina thermal conductivities with respect to temperature in Fig. 5 shows that the TC of the polymer increases with adding the alumina nanoparticle which is in agreement with the experimental results [29]. Despite this consistency between the experimental and the simulated results, some deviations in the value of the TC for the PMMA/alumina nanocomposite is observed which is because of the voids creation due to adding the nanoparticles. Other factors could be related to agglomeration of nanoparticles in particular cites and cavities.

## 4.3 Heat capacity

The specific heat capacities at constant pressure (Cp) or at constant volume (Cv) can be calculated using the fluctuations properties [30] as shown in **Error! Reference source not found.**8 and 9, respectively

$$C_v = \langle \overline{C_v} \rangle = \left\langle \frac{\overline{E^2} - \overline{E}^2}{k_B T^2} \right\rangle \quad (8)$$

$$C_p = \langle \overline{C_p} \rangle = \left\langle \frac{\overline{H^2} - \overline{H}^2}{k_B T^2} \right\rangle \quad (9)$$

where E is the internal energy and H is the enthalpy of the system. The bar sign stands for a time average over a period of time during which one molecular dynamics simulation with the NPT ensemble is performed; and the brackets describe the fact that the average is taken over the 3 optimized configurations for which the simulated Tg has been derived.

As a polymeric system is cooled from the molten state it experiences a different amount of heat capacity before Tg and after it. Thus, Tg can be determined by tracing the change in heat capacity with respect to the temperature. It is notable that the transition happens over a range of temperatures rather than at a single temperature. The temperature in the middle of the aforementioned region is taken as the Tg.

The simulated heat capacities for the PMMA were determined and were shown in Figure 4. As clearly seen, for both the PMMA and the PMMA/Alumina nanocomposite an observable jump appears in the heat capacity around Tg. Despite the good agreement between the experimental and simulated results[31], some deviations are observed. These divergences are higher values of Tg; a broader transition temperature range and a lower difference in the heat capacity between the liquid and glass states compared to the experimental values. Also, comparing the results for the PMMA/alumina and the PMMA shows a lower heat capacity and a lower difference in the heat capacity between the liquid and glass states for the PMMA/Alumina compared to the PMMA system.

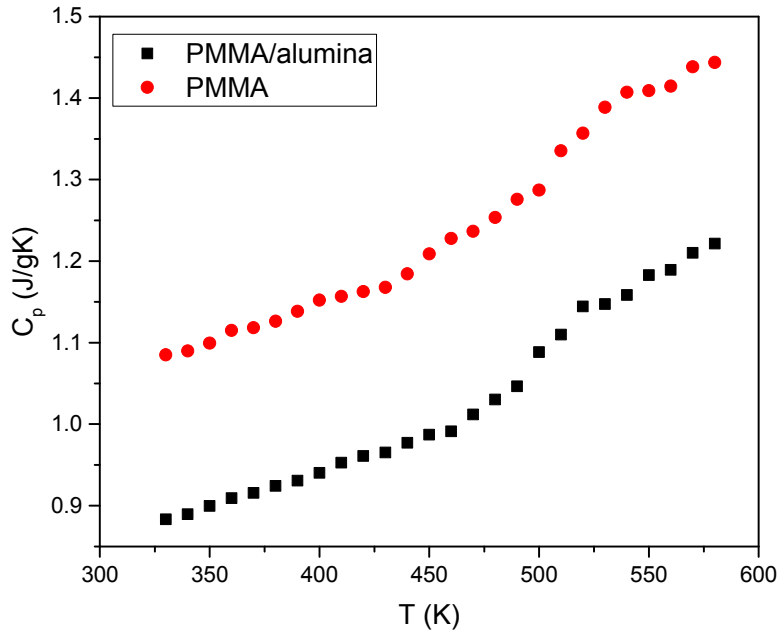

Figure 4. The heat capacity of the PMMA and the PMMA/alumina nanocomposite versus temperature

The jump in the heat capacity which happens at the glass transition could be explained by the requirement of an extra energy to create the volume needed for larger amplitude motions and vibrations [32-34]. Besides, at lower temperatures, vibrational motion practically provides the only contribution to the heat capacity. As the temperature increases, large-amplitude conformational, rotational, and translational motions may also be added to the heat capacity [35].

## 4.4 Thermal diffusivity

In the previous sections, the density, the thermal conductivity and the heat capacity were calculated. Now, according to Eq.1 the thermal diffusivity can be calculated.

Figure 5 indicates the thermal diffusivity for the neat PMMA and the PMMA/Alumina nanocomposite versus temperature .The result reveal that the thermal diffusivity is decreased with increasing temperature and thermal diffusivity of the PMMA/Alumina nanocomposite is greater than thermal diffusivity of the PMMA.The change of the thermal diffusivity during the glass transition is more likely related to the change in the heat capacity and density.

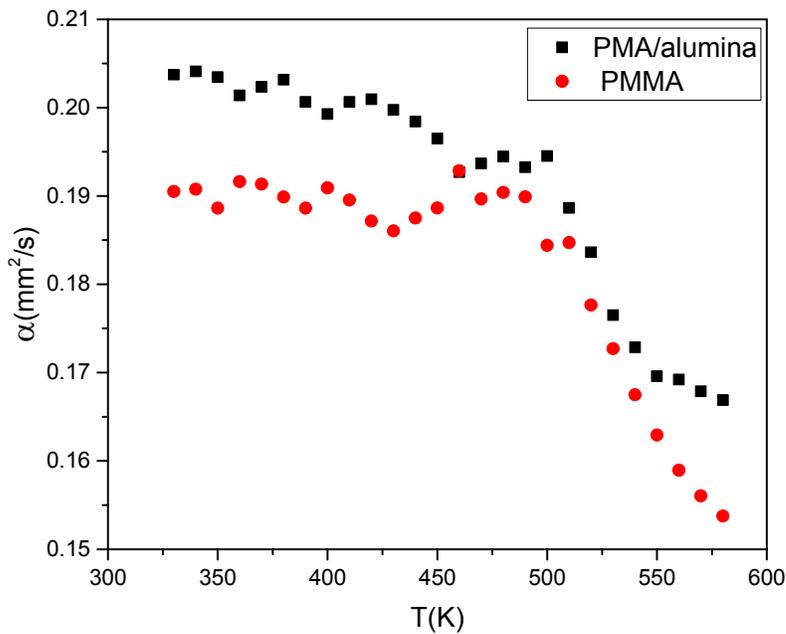

Figure 5.Thermal diffusivity of the PMMA and the PMMA/alumina versus temperature.

## 5   Conclusion

Literature has shown that there exist relationships between the thermal diffusivity and the thermal stability. Also, it can be related to some fire retardant properties such as the total-heat-release (THR), time-to-ignition (TTI) and peak-heat-rare-release (PHRR) that are of the most important parameters in the evaluation of the potential fire hazard of a given material. Metal oxides as one of the most promising flame retardant additives, improve the fire-retardant and the thermal stability properties of polymers. In the present study, MD simulations are used to study the effect of alumina nanoparticles on the thermal diffusivity of the PMMA. The thermal diffusivity of the PMMA and the PMMA/alumina were investigated through calculating heat capacity, density and thermal conductivity in the range of 300-700 K.

The fluctuations properties were employed to calculating the heat capacity. Thermal conductivity was calculated through the nonequilibrium MD (NEMD) simulation by means of the Fourier's law approach. Results show that the alumina nanoparticles decrease the amount of the heat capacity. The heat capacity reduction in the rubbery state (10 percent) is more than in the glassy state (4 percent). The trend of heat capacity versus temperature is in agreement with the experimental results, but the heat capacity amount is less than and the Tg obtained from the heat capacity-temperature curve is greater (about 50 K) than that of the experimental results. The alumina nanoparticles increase the Tg (about 10 K), the thermal conductivity (nearly 10 times) and the thermal diffusivity of the PMMA.

A further study would be useful by investigating other key variables including the cooling rate, polymer molecular weight and the utilization of reactive force field into the simulations for a better scrutinizing of the generalizability and usefulness of the existing MD techniques in determination of thermal properties particularly the thermal diffusivity.

# 6   Acknowledgements

The authors would like to thank HR. Rezaei for his valuable comments and suggestions.

[17]:

$$\alpha = \frac{\kappa}{C_p \rho} \qquad (1)$$

where $\kappa$ is the thermal conductivity, $C_p$ is the heat capacity at a constant pressure, $\rho$ is density and $\alpha$ is the thermal diffusivity. Therefore, in this paper the thermal diffusivity can be achieved by calculating the thermal conductivity, heat capacity and density.

# 8  Materials and MD Simulation Setup

All molecular simulations were done using the MD simulation package, LAMMPS (large-scale atomic/molecular massively parallel simulator), developed at the Sandia National Laboratories [18]. The system consisted of an alumina nanoparticle with the radius of 7 angstroms at the center of the simulation cubic and three linear isotactic PMMA (is-PMMA) chains with a degree of polymerization of 100 around the nanoparticle (10%weight alumina). The structure of the PMMA monomer was proposed by Shaffer et al. [19]. The data of alumina structure is taken from Mincryst [20], the crystallographic database from the Institute of Experimental Mineralogy of the Russian Academy of Sciences.  The PMMA conformations using a random and self-avoiding walk algorithm and the alumina structure were generated in the Fortran 90 and then were imported into the LAMMPS as the data input. The chain structure of the PMMA and the alumina nanoparticle generated as described are shown in Figure 1. To perform the simulations, a three-dimensional periodic boundary condition was applied. To reduce the computation time, a united atom model (where each carbon is grouped with its bonded hydrogen atoms) was applied. An interatomic force field proposed by Okada et al. was used for modeling of the atomic interactions of the polymer chains [21]. Some modifications have been made to the force field equations to make them usable in the LAMMPS. Here, the interaction potential ( $U$ ) is expressed as the sum of the bonding and nonbonding interactions and is defined as the following expression

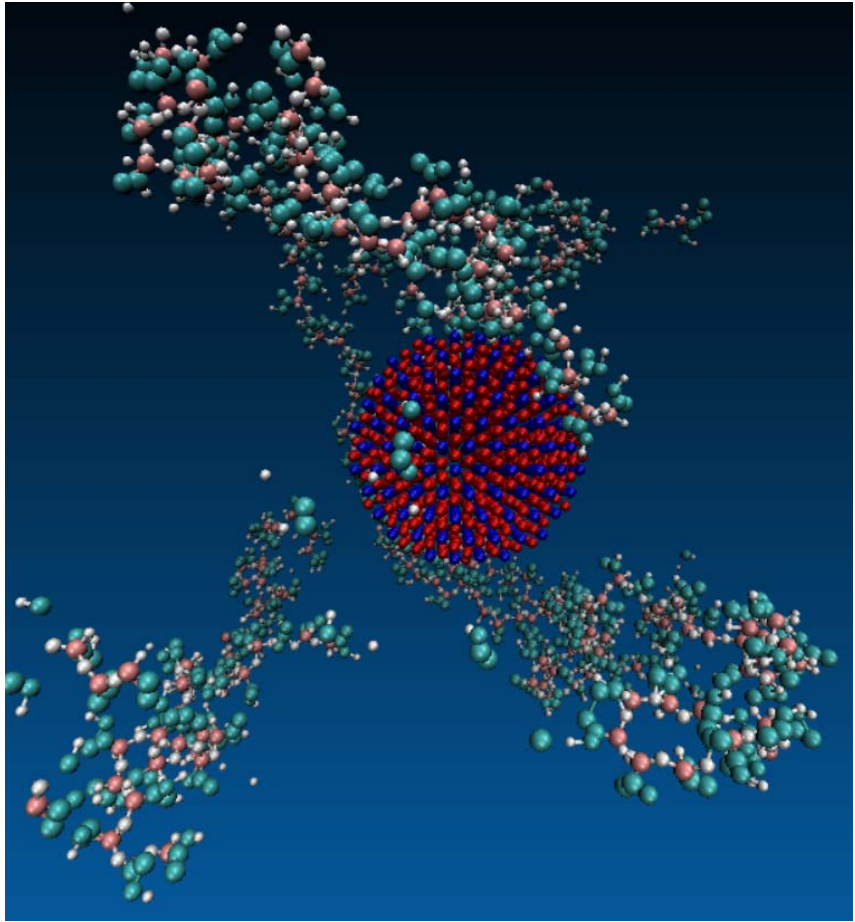

Figure 1. is-PMMA/alumina nanocomposite

$$U = \sum_{bonds} k_r(r-r_0)^2 + \sum_{angles} k_\theta(\theta-\theta_0)^2 + \sum_{torsions}\sum_{i=1}(V_n \cos n\phi)$$
$$+ \sum_{\substack{improper \\ torsions}} (K_1(\Theta-\Theta_0) + K_2(\Theta-\Theta_0)^2) + \sum \frac{A}{r^{12}} - \frac{C}{r^6} \quad (1)$$

where the first term is related the bond stretching energy, $r$ and $r_0$ are the bond length and the equilibrium length of the bond, respectively. The second term illustrates the angular bending energy, $\theta$ and $\theta_0$ are the bending angle and the equilibrium angle of the bond, respectively. The third and fourth terms correspond to the dihedral torsion and the improper torsion energies, respectively. $\phi$ is the dihedral torsion angle and $\Theta$ is the sum of the three neighboring bending angles, $\Theta_0$ is the equilibrium sum of the three neighboring bending angles. The last term is the Lennard–Jones energy between two non-bonded atoms/molecules.

The interaction potential between the PMMA and the alumina in the PMMA/alumina nanocomposite consisted of binding and non-binding potentials [22]. The binding potential defines the interaction between carbonyl and ester oxygen of the PMMA and the aluminum. The non-binding potential describes the interaction between carbons in the PMMA and the aluminum. The binding ($U_b$) and non-binding ($U_{nb}$) potentials are expressed as

$$U_b = A\exp(-2\lambda(z_i - z_b)) - B\exp(-\lambda(z_i - z_b)) + C\lambda^2(z_i - z_b)^2 \exp(-\lambda(z_i - z_b))$$

$$U_{nb} = \frac{2}{3}\pi\sigma_i\varepsilon_i\rho\left[\frac{2}{15}\left(\frac{\sigma_i}{z_i}\right)^9 - \left(\frac{\sigma_i}{z_i}\right)^3 + \frac{2}{3}\sqrt{\frac{5}{2}}\right] \quad (2)$$

where $\rho$ is the number density of aluminum atom, $z_i$ indicates the distance of atom i from the aluminum surface, $\varepsilon_i$ and $\sigma_i$ are Lennard-Jones parameters between the non-binding atoms of the PMMA and the aluminum atoms in the substrate. The first two terms in Equ. (2) are the standard Morse potential and the third term represents the activation energy.

It is notable that since suggested potential by Shaffer and Chakraborty were written for united atom PMMA, there were no potential interactions with hydrogen in PMMA. The potential was expanded to alumina by assuming the interaction of the oxygen atoms in the alumina with PMMA could be described as the average non-binding potential interaction of the aluminum with the PMMA [22]. The First Principles Study on Polymer-Metal-Oxide Adhesion
Was done by Drabold et al. Their result shows the oxygen atoms in alumina have very weak interactions to PMMA and the aluminum atoms interactions with the carbon and hydrogen in PMMA is weak too [23].

The suggested interatomic interactions for the alumina potential proposed by Streitz and Mintmire [24] define a many-body potential consisting of an embedded-atom ($V_{EAM}$) and an electrostatic ($V_{ES}$) part as fallows

$$V_{EAM} = \sum_i F_i[\rho_i] + \sum_{i<j} \varphi_{ij}(r_{ij}) \quad (3)$$

$$\rho_i(r) = \sum_{i\neq j} \xi_j e^{-\beta(r_{ij}-r_j^*)}$$

$$F_i(\rho_i) = -A_i\sqrt{\frac{\rho_i}{\xi_i}}$$

$$\varphi_{ij}(r) = 2B_{ij}e^{\frac{\beta_{ij}}{2}(r-r_{ij}^*)} - C_{ij}\left[1+\alpha(r-r_{ij}^*)\right]e^{-\alpha(r-r_{ij}^*)}$$

Here, $F_i(\rho_i)$ represents the energy required to embed atom $i$ in an environment with an electron density $\rho_i$ and $\varphi_{ij}(r)$ is the pair-wise interaction.

The pair potential $\varphi_{ij}(r)$ becomes strongly repulsive at a small interatomic distance. The electrostatic part accounts for electric charges on atoms and is defined as

$$V_{ES} = \sum_i v_i(q_i) + \frac{1}{2}v_{ij}(r_{ij};q_i;q_j) \qquad (4)$$

$$v_i(q_i) = v_i(0) + \chi_i^0 q_i + \frac{1}{2}J_i^0 q_i^2$$

$$v_{ij}(r_{ij};q_i;q_j) = \int d^3r_1 \int d^3r_2 \rho_i(r_1;q_i)\rho_j(r_2;q_j)/r_{12}$$

## 9 Equilibrium

The three polymer chains were first built up around a spherical cavity with the same diameter of an alumina nanoparticle (For a neat polymer, the diameter was considered zero) in the simulation box at the temperature of T=700K. The polymer chains could move into the empty space where the nanoparticle would be inserted. To avoid this, a wall potential was placed around the space. The wall potential was a Leonard Jones one which parameters were derived from the average Leonard Jones parameters of Okada et al.'s PMMA potential [21]. To minimize the total potential energy of the initial system, the conjugate gradient method was applied.

This was followed by allowing the polymer chains to equilibrate. The equilibration of the polymer has been explained in previous work[25]. Next, the alumina nanoparticle was inserted into the cavity and then the system reached the equilibrium with the NPT ensemble after 2 ns. Then the system was cooled from 700 K to 300 K at a rate of 10K/5ns.

# 10 Results and discussion

## 10.1 Density

Figure 2. illustrates the density ($\rho$) of the PMMA system (neat polymer) and the PMMA/alumina nanocomposite versus a temperature ramp. An obvious change in the slope of the curve is observed. The temperature at which the slope of the curve undergoes a sharp variation is regarded as the glass transition temperature ($T_g$). At $T_g$, the physical properties of a polymer changes and the polymer transfers from a rubbery state (above Tg) to a glassy state (below Tg). According to Figure 2. both the density and the Tg of the PMMA/alumina nanocomposite are larger than those of the PMMA. Figure 2 shows an increase of about 10 K in $T_g$ for the PMMA/alumina nanocomposite compared to that for the PMMA.

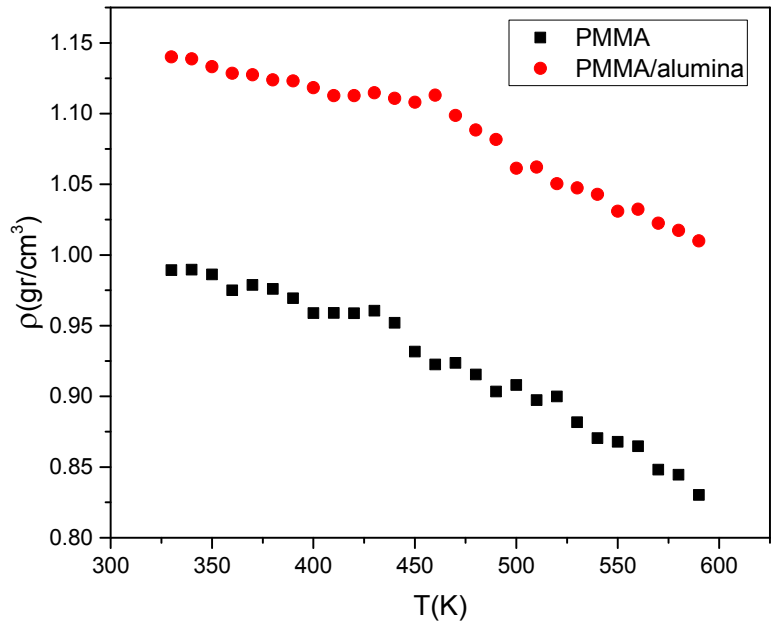

Figure 2. Density of (a) the PMMA and (b) the PMMA/alumina against temperature.

A polymer consists of two phases; one is the polymer chains and the other is the free volume called space phase. According to Figure 2, the density-temperature curve of the PMMA/alumina shows a smaller gradient than that of a pure polymer system, which could be due to a smaller free volume available for the motion of the polymer chains.

## 10.2 Thermal conductivity

There are two methods for the computation of the thermal conductivity (TC) based on the molecular dynamics which are the equilibrium MD (EMD) and the non-equilibrium MD (NEMD) approaches. In this study, the EMD simulation by means of the Fourier's law (**Error! Reference source not found.**6) was employed to investigate the thermal conductivity of the PMMA and alumina/PMMA nanocomposite.

$$\kappa = -\frac{J_q}{dT/dy} \quad (6)$$

where $\kappa$ is the thermal conductivity, dT/dy is the steady-state temperature gradient in unidirectional heat flow inside of the unit cell and $J_q$ is the heat flux that is given with the following expression:

$$J_q = \frac{E}{2A_{cross-section}} \quad (7)$$

The simulation system was divided into several slabs on the total length, two slab at the two end will be the hot region and at the other end will be the cold region. Then the linear temperature region using the least square method was fitted to obtain the temperature gradient. The thermal conductivity κ can be calculated by **Error! Reference source not found.**6.

Figure 3 shows the TC versus temperature for the neat PMMA and the PMMA/alumina nanocomposite. It is found that with increasing the temperature up to the Tg, the thermal conductivity of the PMMA increases, while it decreases above Tg. This is due to the fact that at temperatures above the Tg, the free-volume increases and air has a low thermal conductivity. Such a temperature-dependent behavior of the thermal conductivity is in a good agreement with the experiments [26].

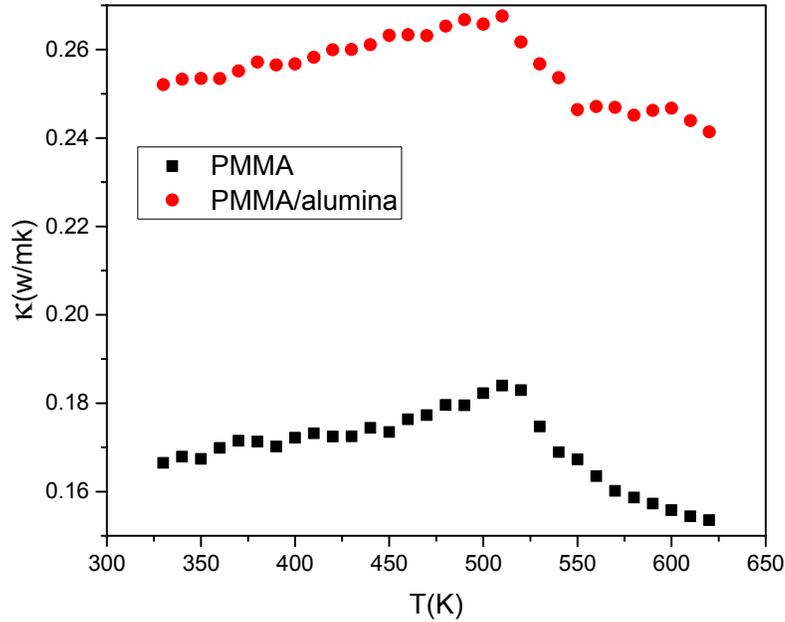

Figure 3. Thermal conductivity of (a) the PMMA and (b) the PMMA/alumina vs. temperature.

Although the thermal conductivity along a single chain is very high [27, 28], thermal conductivity in polymer bulk is very low due to the phonon scattered at and the thermal contact resistance ends of chains. Other limiting factors are random orientation, entanglement chains and voids which act as scattering phonon sites for the heat. The comparison between the PMMA and the PMMA/Alumina thermal conductivities with respect to temperature in Fig. 5 shows that the TC of the polymer increases with adding the alumina nanoparticle which is in agreement with the experimental results [29]. Despite this consistency between the experimental and the simulated results, some deviations in the value of the TC for the PMMA/alumina nanocomposite is observed which is because of the voids creation due to adding the nanoparticles. Other factors could be related to agglomeration of nanoparticles in particular cites and cavities.

## 10.3 Heat capacity

The specific heat capacities at constant pressure ($C_p$) or at constant volume ($C_v$) can be calculated using the fluctuations properties [30] as shown in **Error! Reference source not found.**8 and 9, respectively

$$C_v = \langle \overline{C_v} \rangle = \left\langle \frac{\overline{E^2} - \overline{E}^2}{k_B T^2} \right\rangle \quad (8)$$

$$C_p = \langle \overline{C_p} \rangle = \left\langle \frac{\overline{H^2} - \overline{H}^2}{k_B T^2} \right\rangle \quad (9)$$

where E is the internal energy and H is the enthalpy of the system. The bar sign stands for a time average over a period of time during which one molecular dynamics simulation with the NPT ensemble is performed; and the brackets describe the fact that the average is taken over the 3 optimized configurations for which the simulated $T_g$ has been derived.

As a polymeric system is cooled from the molten state it experiences a different amount of heat capacity before $T_g$ and after it. Thus, $T_g$ can be determined by tracing the change in heat capacity with respect to the temperature. It is notable that the transition happens over a range of temperatures rather than at a single temperature. The temperature in the middle of the aforementioned region is taken as the $T_g$.

The simulated heat capacities for the PMMA were determined and were shown in Figure 4. As clearly seen, for both the PMMA and the PMMA/Alumina nanocomposite an observable jump appears in the heat capacity around $T_g$. Despite the good agreement between the experimental and simulated results[31], some deviations are observed. These divergences are higher values of $T_g$; a broader transition temperature range and a lower difference in the heat capacity between the liquid and glass states compared to the experimental values. Also, comparing the results for the PMMA/alumina and the PMMA shows a lower heat capacity and a lower difference in the heat capacity between the liquid and glass states for the PMMA/Alumina compared to the PMMA system.

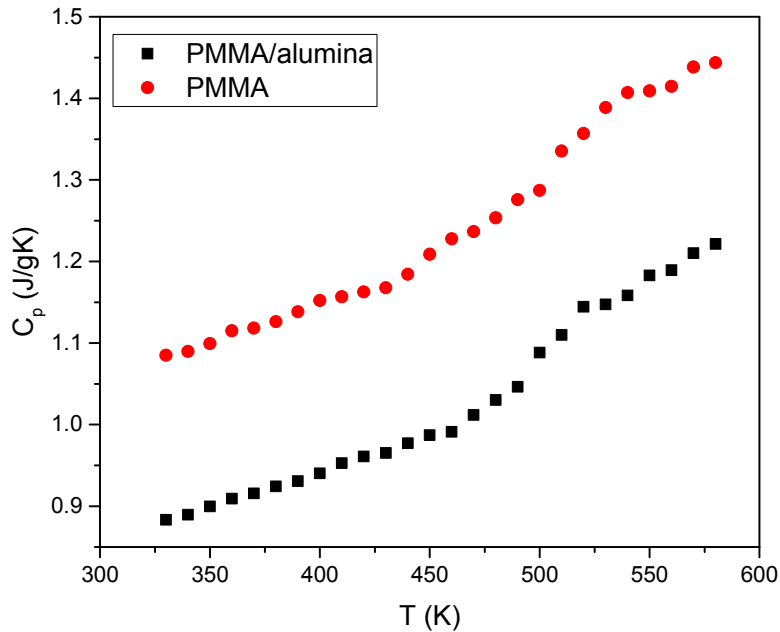

Figure 4. The heat capacity of the PMMA and the PMMA/alumina nanocomposite versus temperature

The jump in the heat capacity which happens at the glass transition could be explained by the requirement of an extra energy to create the volume needed for larger amplitude motions and vibrations [32-34]. Besides, at lower temperatures, vibrational motion practically provides the only contribution to the heat capacity. As the temperature increases, large-amplitude conformational, rotational, and translational motions may also be added to the heat capacity [35].

## 10.4 Thermal diffusivity

In the previous sections, the density, the thermal conductivity and the heat capacity were calculated. Now, according to Eq.1 the thermal diffusivity can be calculated.

Figure 5 indicates the thermal diffusivity for the neat PMMA and the PMMA/Alumina nanocomposite versus temperature .The result reveal that the thermal diffusivity is decreased with increasing temperature and thermal diffusivity of the PMMA/Alumina nanocomposite is greater than thermal diffusivity of the PMMA.The change of the thermal diffusivity during the glass transition is more likely related to the change in the heat capacity and density.

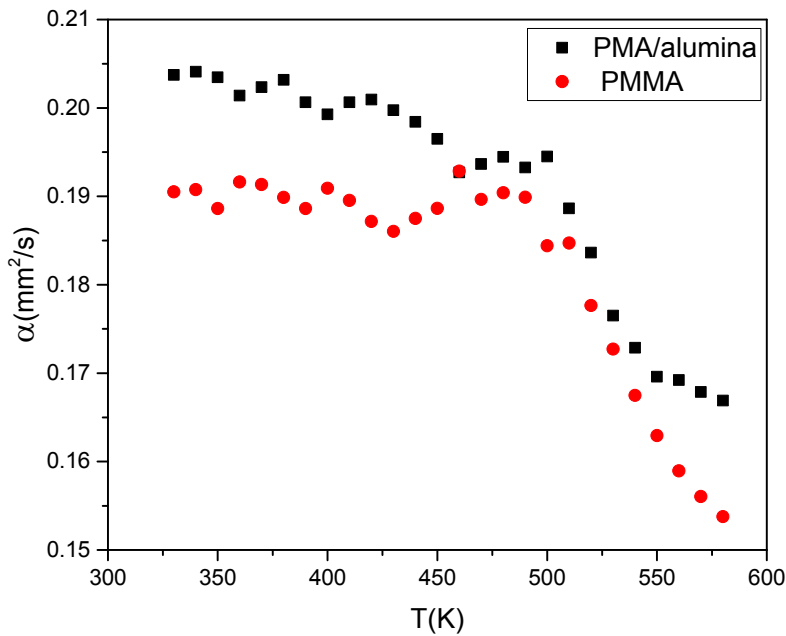

Figure 5.Thermal diffusivity of the PMMA and the PMMA/alumina versus temperature.

## 11 Conclusion

Literature has shown that there exist relationships between the thermal diffusivity and the thermal stability. Also, it can be related to some fire retardant properties such as the total-heat-release (THR), time-to-ignition (TTI) and peak-heat-rare-release (PHRR) that are of the most important parameters in the evaluation of the potential fire hazard of a given material. Metal oxides as one of the most promising flame retardant additives, improve the fire-retardant and the thermal stability properties of polymers. In the present study, MD simulations are used to study the effect of alumina nanoparticles on the thermal diffusivity of the PMMA. The thermal diffusivity of the PMMA and the PMMA/alumina were investigated through calculating heat capacity, density and thermal conductivity in the range of 300-700 K.

The fluctuations properties were employed to calculating the heat capacity. Thermal conductivity was calculated through the nonequilibrium MD (NEMD) simulation by means of the Fourier's law approach. Results show that the alumina nanoparticles decrease the amount of the heat capacity. The heat capacity reduction in the rubbery state (10 percent) is more than in the glassy state (4 percent). The trend of heat capacity versus temperature is in agreement with the experimental results, but the heat capacity amount is less than and the $T_g$ obtained from the heat capacity-temperature curve is greater (about 50 K) than that of the experimental results. The alumina nanoparticles increase the Tg (about 10 K), the thermal conductivity (nearly 10 times) and the thermal diffusivity of the PMMA.

A further study would be useful by investigating other key variables including the cooling rate, polymer molecular weight and the utilization of reactive force field into the simulations for a better scrutinizing of the generalizability and usefulness of the existing MD techniques in determination of thermal properties particularly the thermal diffusivity.

## 12  Acknowledgements

The authors would like to thank HR. Rezaei for his valuable comments and suggestions.